\documentclass[aps,prb,floatfix,twocolumn,reprint,superscriptaddress]{revtex4-1}
\usepackage{amsfonts}
\usepackage{mathrsfs}
\usepackage{amsmath}% needed for subequations
\usepackage{color}
\usepackage{graphicx}
\usepackage{bm}% bold maths
\usepackage{amssymb}
\usepackage{xspace}
\usepackage{epstopdf}
\usepackage{dcolumn}% Align table columns on decimal point
\usepackage{longtable}
\usepackage{multirow}
\usepackage[colorlinks=true, letterpaper=true, pdfstartview=FitV, linkcolor=blue, citecolor=blue, urlcolor=blue]{hyperref}

\makeatletter

\newcommand{\Rmnum}[1]{\expandafter\@slowromancap\romannumeral #1@}
\makeatother

\begin{document}

\title{Ultra-Stable Ferrimagnetic Second-Order Topological Insulator in 2D Metal-Organic Framework}

\author{Meijun Wang}
\author{Yong-An Zhong}
\author{Lei Jin}\email{jinlei994@163.com}
\author{Ying Liu}
\author{Xuefang Dai}
\author{Guodong Liu}
\author{Xiaoming Zhang}\email{zhangxiaoming87@hebut.edu.cn}

\affiliation{ State Key Laboratory of Reliability and Intelligence of Electrical Equipment, Hebei University of Technology, Tianjin 300130, China.}

\affiliation{ School of Materials Science and Engineering, Hebei University of Technology, Tianjin 300130, China.}

%\date{\today}
\begin{abstract}
Two-dimensional (2D) magnetic second-order topological insulators (SOTIs) exhibit distinct topological phases characterized by spin-polarized zero-dimensional (0D) corner states, which have garnered significant interest. However, 2D ferrimagnetic (FiM) SOTIs, particularly those that simultaneously exhibit ultra-stable corner states, are still lacking. Here, based on first-principles calculations and theoretical analysis, we reveal such SOTI state in a 2D metal-organic framework (MOF) material, Cr(pyz)$_2$ (pyz = pyrazine). This material exhibits FiM ground state with an easy axis aligned along [001] direction. It hosts a nontrivial real Chern number in the spin-up channel, enabled by $PT$ symmetry, with 0D corner states observable in disk. In contrast, the spin-down channel exhibits a trivial gapped bulk state. Notably, the topological corner states in monolayer Cr(pyz)$_2$ show high robustness, even if the symmetries are broken by introducing defects, the corner states persist. We also considered other external perturbations, including uniaxial/biaxial strain, ligand rotation, and electric fields, the corner states still remain stable. Even more, the energy positions of the corner states are also nearly unchanged. This work is the first to identify ultra-stable FiM SOTI state in the MOF system, and provide an ideal platform for future experimental investigations and applications in spintronic devices.

\end{abstract}

\maketitle

\section{Introduction}
The identification of topological insulators (TIs) has ushered in a new era in condensed matter physics and materials science~\cite{hasan2010colloquium,qi2011topological,zhang2009topological,liu2010model,xia2009observation,hsieh2009tunable}. A key characteristic of TIs is the bulk-boundary correspondence, that is conventional d-dimensional TIs host robust gapless states on their $d-1$ dimensional boundaries. For example, three-dimensional (3D) TIs exhibit 2D surface states, while 2D TIs support one-dimensional (1D) edge states. Subsequent research revealed that robust gapless states can also emerge on $d-2$ dimensional boundaries, thereby putting forward the SOTI concept~\cite{sheng2019two1,chen2021graphyne,qian2021second,pan2022two}. Taking 2D SOTIs as examples, they feature 0D corner states, while their 1D edge states open gap. Depending on the different symmetry protection mechanisms of the corner states, these 2D SOTIs can be characterized by real Chern number ($v_R$) or the quantized fractional topological charge ($Q_c$)~\cite{chen2021graphyne,benalcazar2019quantization,takahashi2021general}.

To date, SOTIs states have been identified in various 3D and 2D materials, including SnTe~\cite{schindler2018higher}, Bi$_4$Br$_4$~\cite{zhao2023topological}, graphdiyne~\cite{sheng2019two2}, BiSe~\cite{costa2021discovery}, and the MOF material Ni$_3$(C$_6$S$_6$)$_2$~\cite{he2023metal}. Researchers have also investigated the effects of external perturbations on these states. For instance, the topological transition occurs between SOTIs and topological crystalline insulators (TCIs) under magnetic field~\cite{ezawa2018topological}. Furthermore, applying electric field can break the degeneracy of corner states in bismuthine, causing their energy levels to oscillate with the field direction~\cite{yan2022plane}. In layered CrOCl, the vertical electric field can induce a phase transition from high-order topological insulator to \text{``}hidden higher-order topological insulator\text{''} phase, submerging the corner states into edge or bulk states~\cite{guo2023magnetic}. However, considering the special device applications in certain scenarios, stable SOTIs are essential. The stable SOTI states are also more conducive to experimental observation. Therefore, the search for stable SOTI is necessary.

In addition to investigating the intrinsic properties of SOTIs, the combination of SOTIs with magnetism is also an important research trend, which has led to the realization of various magnetic SOTI phases. Currently, SOTI states have been observed in magnetic materials EuIn$_2$As$_2$~\cite{xu2019higher}, bismuthene on EuO (111) surface~\cite{chen2020universal}, NpSb~\cite{mao2021magnetism}, 2H-RuCl$_2$~\cite{sheng2022strain}, CrOCl~\cite{guo2023magnetic}, CrSiTe$_3$~\cite{wang2023magnetic}, FeSe~\cite{mu2022antiferromagnetic}, 2H-VX$_2$ (X = S, Se, Te)~\cite{liu2023magnetic}, MOF material Co$_3$(HITP)$_2$~\cite{zhang2023magnetic}, and others. However, these magnetic SOTI states are usuaiiy identified in ferromagnetic (FM) or antiferromagnetic (AFM) materials Notably, FiM system combines the advantages of the easy detection of FM materials and the ultrafast magnetic dynamics of AFM materials, which have many potential application, such as computing in memory, neural network device, skyrmions logic gate, FiM tunnel junction~\cite{dong2021electrically, wang2021ultrafast, zhang2019skyrmion, reza2019fast, zhang2023ferrimagnets}. The combination of SOTIs with ferrimagnetism is expected to give rise to the new topological phases and more spin-related functional characteristics. Nevertheless, the FiM SOTIs remain rare, and their realistic materials, underlying mechanisms and potential application still require further exploration and development.

In this work, we discovered such material, namely Cr(pyz)$_2$, which not only is a 2D FiM SOTI, but also has ultra-stable topological corner state. Our results show that the antiparallel arrangement of Cr atom spins relative to the pyz molecules establishes the FiM ordering within the crystal structure. First-principles calculations confirm that both spin channels of the material exhibit insulating behavior and have large energy gaps. The spin-up channel shows a nontrivial gapped bulk state with real Chern number of 1, while the spin-down channel has trivial gap. According to second-order bulk-boundary correspondence, the material possesses topologically protected corner states in square disk. Notably, these corner states exhibit robustness under various conditions, including strain, ligand rotation, and electric fields. Even with symmetry-breaking perturbations, such as introducing holes at the edges and inside of the disk, the corner states remain stable. These results indicate that the FiM Cr(pyz)$_2$ monolayer has ultra-stable corner states. Our work proposed a new type of spin-polarized SOTI with robust topological corner states in MOF system, which is promising to be observed experimentally in the future and has potential application value in the field of spintronics.

\section{Computational Methods}
The first-principles calculations in this study were performed using the Vienna ab initio simulation package (VASP), within the framework of density functional theory (DFT)~\cite{kresse1996efficiency1, kresse1996efficient2}. The exchange correlation effects were treated with the generalized gradient approximation (GGA) using the Perdew-Burke-Ernzerhof (PBE) functional~\cite{perdew1996generalized}. To accurately describe the ionic potentials, the projector augmented wave (PAW) approach was employed. A vacuum space of 20\AA was introduced in the crystal structure to avoid potential interactions between layers. The cutoff energy was set as 500 eV. The Brillouin zone was sampled with a $5\times5\times1$ Monkhorst-Pack $k$-point grid~\cite{monkhorst1976special}.  To account for Coulomb interactions, we utilized the DFT+U method during our calculations~\cite{yang1999influence, anisimov1991band}. The effective $U$ value for Cr was chosen as 3 eV~\cite{li2021two}. The lattice parameters and ionic positions were optimized until the residual force on each atom is less than 0.01 eV/\AA~. The convergence criterion for energy was set to $10^{-6}$  eV. The symmetry character of the electronic states was analyzed using the Irvsp code~\cite{gao2021irvsp}. To investigate the corner modes, an ab initio tight binding model was constructed using the Wannier90 package~\cite{pizzi2020wannier90}, followed by disk spectrum calculations using the pybinding package~\cite{moldovan2017peeters}.

\section{Results}
\begin{figure}
  \centering
  \includegraphics[width=8.5cm]{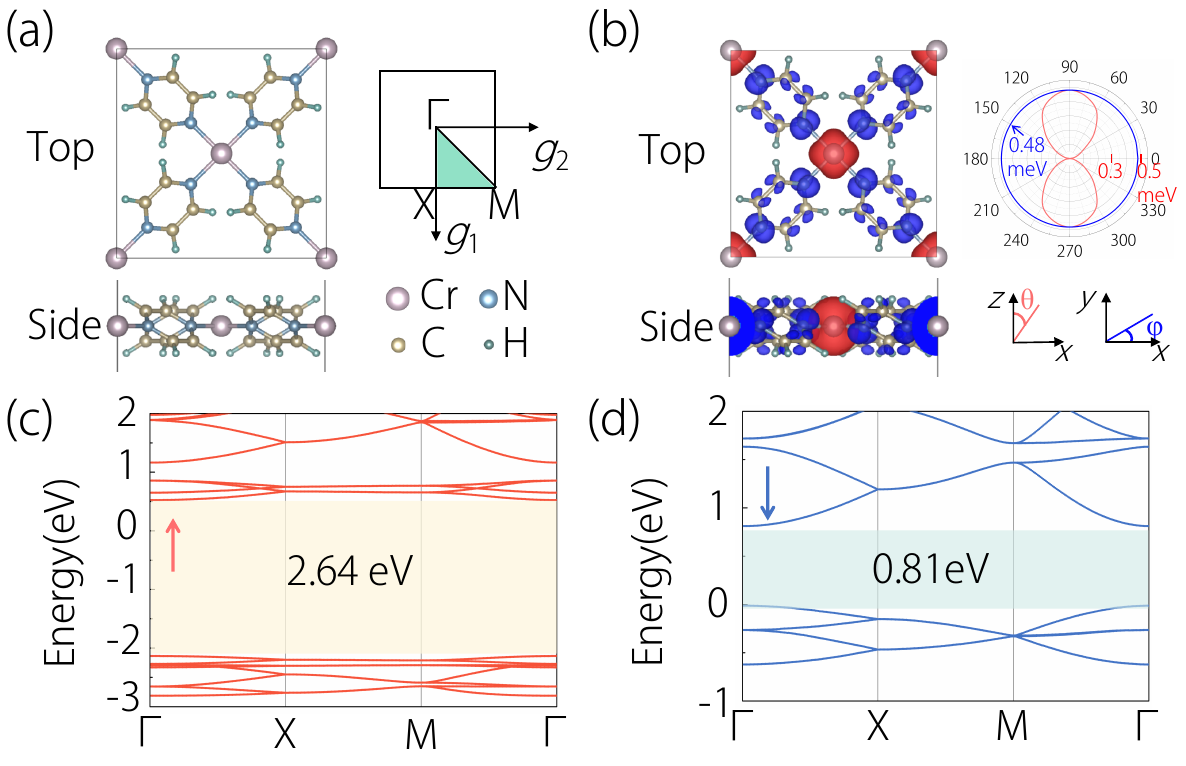}
  \caption{(a) Top and side views of the crystal structure of the Cr(pyz)$_2$ monolayer, and corresponding the 2D Brillouin zone. (b) The spin density distribution of FiM state for Cr(pyz)$_2$ monolayer, and the MAE upon rotating the spin within the $z-x$ and $y-x$ planes. The red and blue lines represent the fitted curves; $ MAE_{(\theta or\varphi)}=E_{(\theta or\varphi)} -E_{(\theta =0)} $ (where $\theta$ and $\varphi$ are the polar and azimuthal angles, respectively). (c) and (d) are the spin-up and spin-down band structures of Cr(pyz)$_2$ monolayer.}\label{figure1}
\end{figure}

The lattice structure of Cr(pyz)$_2$ is illustrated in Fig.~\ref{figure1}(a). This structure belongs to the space group $P4/nbm$ (No.125), with the primitive cell comprising two Cr ions and four pyz molecules~\cite{yang2022two}. In the Cr(pyz)$_2$ layer, each Cr ion is surrounded by four pyz units, forming a coordination pattern that is nearly square planar, while each pyz molecule is linearly coordinated to two neighboring Cr ions. The lattice constant for monolayer Cr(pyz)$_2$ is optimized as $a$ = $b$ = 9.899 \AA,  which align well with previously reported values~\cite{lv2022enhanced}.

The Cr(pyz)$_2$ monolayer is expected to be exfoliated from the experimentally synthesized bulk crystal Li$_{0.7}$[Cr(pyz)$_{2}$]Cl$_{0.7}\cdot$ 0.25(THF) (pyz = pyrazine, THF = tetrahydrofuran)~\cite{perlepe2020metal}. The stability of monolayer Cr(pyz)$_2$ can be evaluated through $ab initio$  molecular dynamics (AIMD) simulations. The spin-polarized AIMD simulation is performed in a $2\times2\times1$ supercell at 300 K, as shown in Fig.S1 of $Supporting Information$. The results demonstrate that no bond breakages or geometric reconfigurations occurred after 3 ps simulation, thereby confirming the thermal stability of monolayer Cr(pyz)$_2$.

The magnetism in monolayer Cr(pyz)$_2$ primarily arises from Cr atoms, as the partially filled $d$-orbitals of transition metal atoms typically contribute to magnetism. In this material, in addition to the Cr atom contributing 3.4 \(\mu_B\), the pyz molecule also contributes 0.6 \(\mu_B\). Moreover, the spin directions of pyz molecules are aligned antiparallel to those of Cr atoms, leading to FiM ground state. This can be evidenced by the spin density shown in Fig.~\ref{figure1}(b). We also compared the energies of FM, AFM, and FiM states to further confirm the magnetic ground state. The FiM state has the lowest energy, which is 1.116 and 0.623 eV lower than that of FM and AFM, respectively. Therefore, monolayer Cr(pyz)$_2$ has a FiM ground state. To determine the easy axis in the FiM state, we calculate the magnetic anisotropy energy (MAE) by rotating the spin within the $z-x$ and $y-x$ planes, using DFT+U+SOC method. The results, displayed in Fig.~\ref{figure1}(b), show that the lowest energy is along the [001] direction. The maximal values of energy difference between [001] direction and the magnetization in the $z-x$ and $y-x$ plane both are 0.48 meV per cell. Therefore, the easy axis of monolayer Cr(pyz)$_2$ in the FiM ground state is along [001], consistent with previous reports~\cite{lv2022enhanced}.

In the FiM state, we calculated the spin-resolved band structures of monolayer Cr(pyz)$_2$ in the absence of SOC, as shown in Figs.~\ref{figure1}(c) and (d). Both the spin-up (red lines) and spin-down (blue lines) band structures exhibit insulating behavior, with band gaps of 2.64 eV and 0.81 eV, respectively. The valence band maximum (VBM) and conduction band minimum (CBM) states belong to different spin channels, resulting in a direct band gap of 0.536 eV. The projected density of states (PDOS) analysis reveals that the spin-up bands near the Fermi level are primarily contributed by the Cr-$d$ and N-$p$ orbitals, while the spin-down bands are mainly contributed by the C-$p$ and N-$p$ orbitals, as shown in Fig.S2. We also examined the effect of different U values on the band structure of monolayer Cr(pyz)$_2$. Figs.S3(a)-(e) presents the band structures with $U$ values of 0, 2, 4, 5 and 6 eV, respectively. The results show that the overall band structure remains unchanged, with only the band gap gradually increasing as the $U$ value rises [see Fig.S3(f)].

In the absence of SOC, the spin channels become independent, allowing each to be treated as an effective spinless system. These spinless systems, i.e., the original ones without magnetism, remain all the crystal symmetries, including time reversal symmetry ($T$). Namely, the system is composed of two spinless subsystem ($\sigma$ =$\uparrow$, $\downarrow$). Each subsystem individually preserves both $T$ and spatial inversion ($P$) symmetries and $(PT)^{2} =1$. When $PT$ symmetry is present, the topology of the system can be described by a real Chern number ($v_{R}^{\sigma} =0, 1$) as the topological invarian. This $v_{R}^{\sigma}$ can be determined using the parity eigenvalues of the valence band at the four inverse invariant momentum points $\Gamma _{i} (i=1, 2, 3, 4)$. The formula for this is:
  $$(-1)^{v_{R}^{\sigma }}=\prod_{i=1}^{4}(-1)^{[n_{i,-}^{\sigma }/2]}$$
where [$\cdot \cdot \cdot$] is the floor function, and $n_{i,-}^{\sigma }$ represents the number of valence states at $\Gamma _{i}$ which have negative $P$ eigenvalue. The value of real Chern number can only be 0 or 1. When it equals 1, it indicates a nontrivial bulk topology with 0D corner states at the sample's corners. The results of $n_{i,-}^{\sigma }$ and $v_{R}^{\sigma}$ in monolayer Cr(pyz)$_2$ have been summarized in Table~\ref{Table1}. We can find that the spin-up channel has nontrivial $v_{R}^{\uparrow}$ =1, whereas the spin-down channel is trivial with $v_{R}^{\downarrow}$ =0. Therefore, the monolayer Cr(pyz)$_2$ belongs to magnetic SOTI in the absence of SOC.
\begin{table*}[t!]
  \centering
  \caption{$P$ eigenvalues and Chern number ($v_{R}$) for both
  spin channels}\label{Table1}
\renewcommand{\arraystretch}{1.5}
\setlength{\tabcolsep}{8mm}{
\begin{tabular}{cccccc}
  \hline
  state & $\Gamma$ & X & M & Y & $v_{R}$ \\
  \hline
  Spin-up& 35& 37& 37& 37& 1 \\
  Spin-down& 35& 35& 35& 35& 0 \\
  \hline
\end{tabular}}
\end{table*}

\begin{figure}
  \centering
  \includegraphics[width=8.5cm]{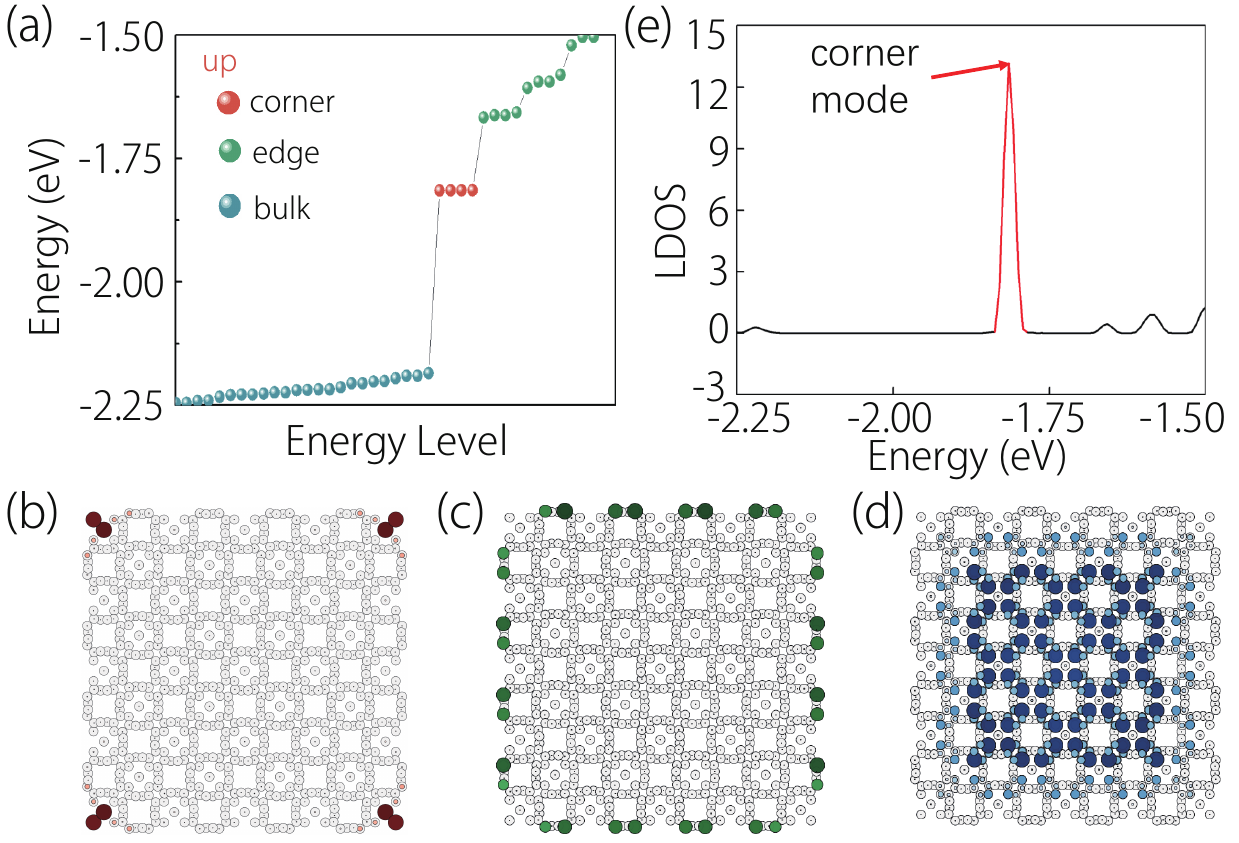}
  \caption{(a) shows the energy spectrum for the Cr(pyz)$_2$ square disk. (b) - (d) The spatial distribution of the corner, edge, and bulk modes for disks, corresponding to the red, green, and blue dots in (a), respectively. (e) The local density of states at one corner of the disk. The sharp peak corresponds to the corner mode.}\label{figure2}
\end{figure}

When the 2D insulator has a real Chern number of \text{``}1{''}, the system will exhibit 0D corner states. As an important symbol of 2D SOTI, corner states can be identified from the edge and bulk states of the system. In our work, the monolayer Cr(pyz)$_2$ has the real Chern number of \text{``}1{''} in spin-up channel, revealing the existence of 0D corner states in the spin-up band gap. To prove this, we first calculated the projected spectra at the edge for spin-up channel, as shown in Fig.S4(a). The presence of band gap can be clearly observed in the projected spectra, which suggests that the material may be different from conventional insulators. By calculating the energy spectrum for the square disk, we indeed observe four degenerate corner states at -1.815 eV, as shown in Fig.~\ref{figure2}(a), which is consistent with the real Chern number results. Further, mapping their positions in real space, these corner states are clearly localized at the four corners of the sample, as illustrated in Fig.~\ref{figure2}(b), directly confirming that monolayer Cr(pyz)$_2$ is a 2D SOTI. Here, the disk has a side length of 4$a$ ($\sim$3.96 nm), where a represents the lattice constant for monolayer Cr(pyz)$_2$. The green and blue balls in Fig.~\ref{figure2}(a) correspond to the edge and bulk states of the material, respectively. For comparison with the corner states, we plot the distribution of the edge and bulk states in real space, as shown in Figs.~\ref{figure2}(c) and (d). In addition, we calculated the local density of states (LDOS) at one corner of the disk [see Fig.~\ref{figure2}(e)], with the mode weight taken within a unit cell at the corner. A sharp peak is observed in the bulk gap, corresponding to the corner modes.

In contrast, the energy band gap for the spin-down channel contains only edge and bulk states, with no corner states, which is consistent with the results obtained from the validated formula above. Fig.S4(b) display the projected spectra at edge for spin-down channel. Although the projected spectra show the existence of energy gaps, there are no points corresponding to corner states in the energy spectrum [see Fig.S4(c)], only edge and bulk states. The spatial locations of edge and bulk modes at different energy levels are illustrated in Fig.S4(d) and (e).

The corner states in monolayer Cr(pyz)$_2$ exhibit remarkable robustness, maintaining their stability even under conditions of broken symmetries. Our work has demonstrated this by introducing defects, applying uniaxial/biaxial strains, rotating ligands and applying electric fields. Next, we will discuss each of them.

\begin{figure}
  \centering
  \includegraphics[width=8.5cm]{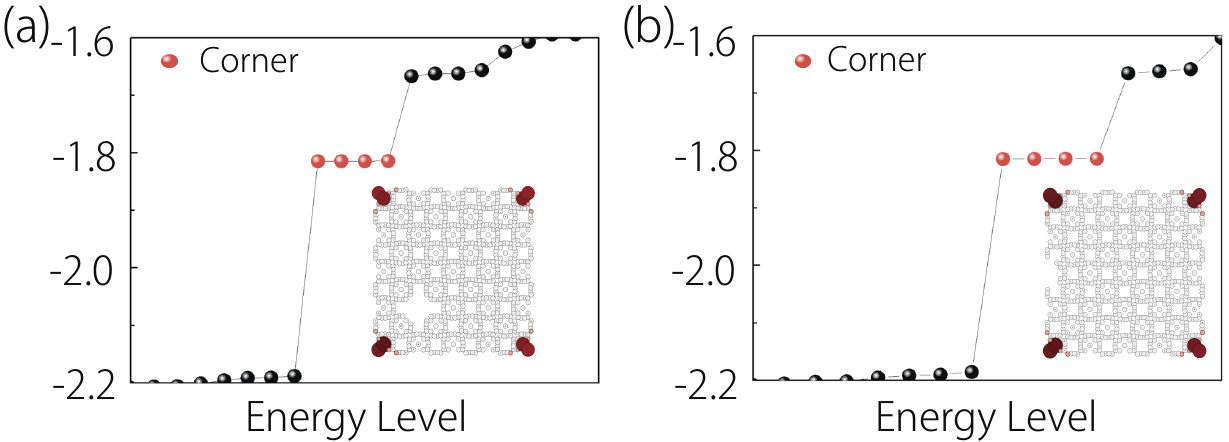}
  \caption{The energy spectrum of spin-up and the distribution of corner modes in real-space were calculated by artificially introducing holes at (a) inside and (b) edge of square shape nano disk. The red balls in the energy spectrum indicate the locations of the corner modes.}\label{figure3}
\end{figure}

We first break symmetries of Cr(pyz)$_2$ to verify the stability of its corner states. As shown in the insets of Figs.~\ref{figure3}(a) and (b), we artificially introduced distortions at the inside or edge of the square disks to break the P symmetry. The results indicate that the corner states of Cr(pyz)$_2$ exhibit exceptionally high stability in the edge and bulk gaps, unaffected by these perturbations. This observation is also consistent with the edge picture perspective: as long as the bulk and edge gaps are preserved, the topological classification for each edge remains unchanged despite the presence of symmetry-breaking perturbations.

\begin{figure}
  \centering
  \includegraphics[width=8.5cm]{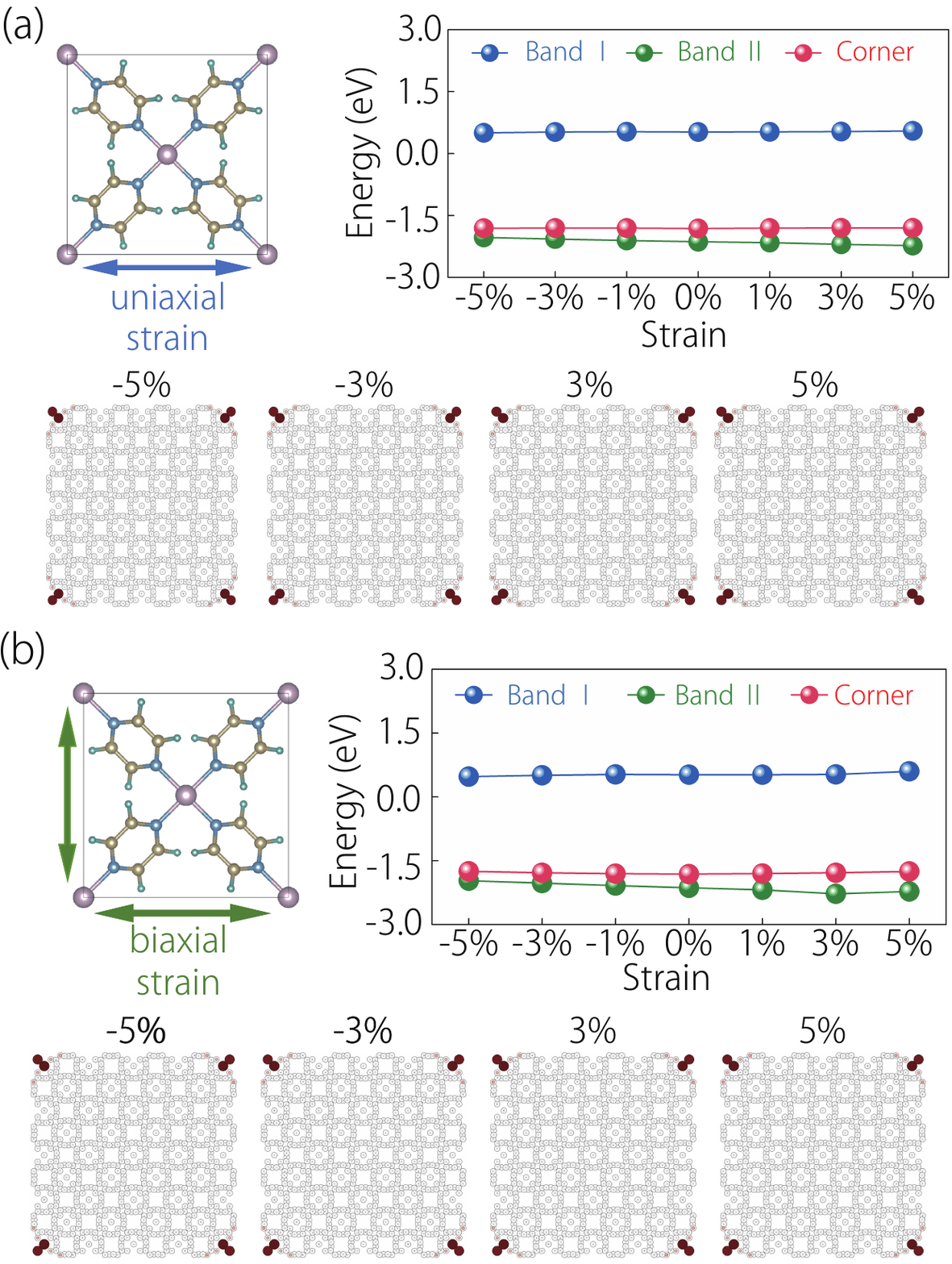}
  \caption{Illustration of the positions of conduction band minimum (CBM), valence band maximum (VBM) and corner state in the spin-up channel under (a) uniaxial strain and (b) biaxial strain. The positive and negative horizontal coordinates represent tensile and compressive strain. The (a) and (b) also display the spatial distribution of corner modes at disks under some strain value.}\label{figure4}
\end{figure}

Next, we take two types of strains on material: uniaxial and biaxial strain. We first consider uniaxial strain effect. Applying tensile and compressive strain along the $a/b$ axis. To ensure the SOTI state, two crucial conditions need to be satisfied: (i) an insulating band gap in the spin-up channel; (ii) the corner states in the band gap. Fig.~\ref{figure4}(a) show the positions of conduction band minimum (CBM) and valence band maximum (VBM) within $\pm5\%$ strain. Here, $\text{``}+\text{''}$ represents tensile strain, and $\text{``}-\text{''}$ denotes the compressive strain. The CBM rise to higher energy level with strain turning from the compressive to tensile one, while VBM rise to lower energy level. As a result, the insulating gap of the spin-up channel corresponding to the Fermi level remains throughout the process and slightly increases. We also check the corner states in monolayer Cr(pyz)$_2$, founding that these corner states always preserved and their energies do not change much. The Fig.~\ref{figure4}(a) also displays the corner state in the square disk under $-5\%, -3\%, 3\%, 5\%$ uniaxial strains.

When biaxial strain is applied, the system exhibits behavior similar to that observed under uniaxial strain. Under biaxial strain, all crystalline symmetries are preserved. Fig.~\ref{figure4}(b) displays that the band gap is always present under strain. As the strain transitions from compression to tension, the energy separation between CBM and VBM gradually increases, resulting in a larger band gap, just the change is small. The energy of corner state is located between CBM and VBM. With the change of strain, the corner state remains consistently present, and its energy position is almost constant. Fig.~\ref{figure4}(b) also illustrates the real-space distribution of the corresponding corner states under various strain conditions. These results indicate that corner states in monolayer Cr(pyz)$_2$ are not affected by uniaxial and biaxial strains.

\begin{figure}
  \centering
  \includegraphics[width=8.5cm]{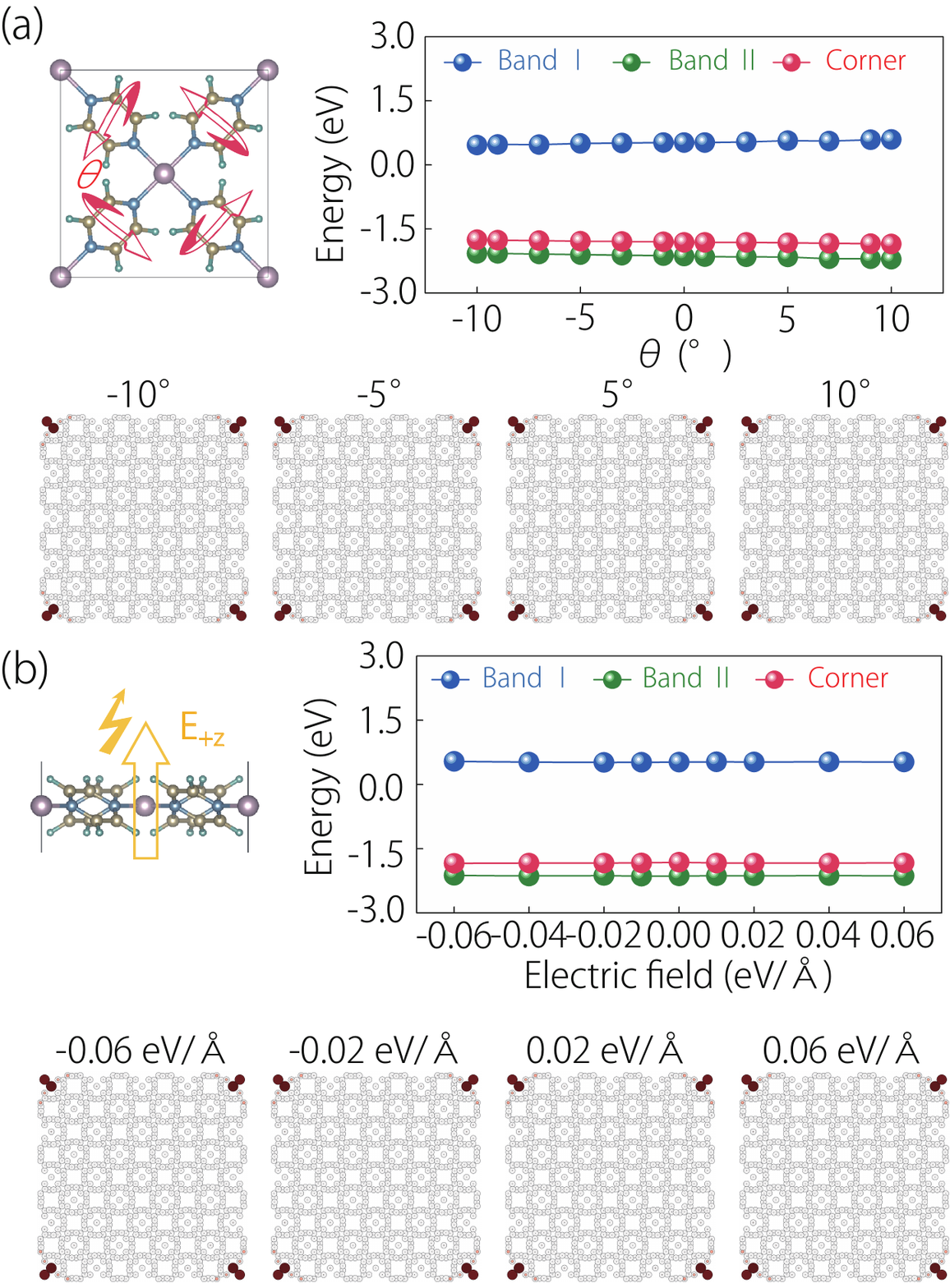}
  \caption{Illustration of the positions of CBM, VBM and corner state in the spin-up channel under (a) rotating ligand and (b) electric field. The positive and negative horizontal coordinates represent clockwise and counterclockwise rotation in (a), and the direction of electric field in (b). The (a) and (b) also display the spatial distribution of corner modes at some ligand rotation angles or electric field strengths.}\label{figure5}
\end{figure}

The monolayer Cr(pyz)$_2$ consists of Cr metal ions and pyz ligands, which are linked together through Cr-N bonds. Here, we discuss the impact of ligand rotation on topological corner states in monolayer Cr(pyz)$_2$. The ligands rotate around the Cr-N bonds. Fig.~\ref{figure5}(a) shows the changes in the energy levels of CBM, VBM and corner states when the ligand is rotated at different angles ($\theta$). The \text{``}$\theta${''} represents the dihedral angle between the pyz ligand and ab lattice plane. The positive \text{``}$\theta${''} represents clockwise rotation of ligands, while negative \text{``}$\theta${''} indicates counterclockwise rotation. The angle \text{``}0{''} corresponds to the intrinsic state of material. As shown in Fig.~\ref{figure5}(a), the energy levels of CBM and VBM, which make up the energy gap, change very little as the ligand rotates from $-10^{\circ}$ to $10^{\circ}$. The maximum energy change of VBM does not exceed 0.136 eV, while that of CBM is even smaller, not exceeding 0.132 eV. Overall, the band gap is consistently present near the Fermi level, which suggests the potential emergence of corner states. By calculating the real Chern numbers and energy spectrum, we confirm that the material still hosts corner states when the dihedral angle $\theta$ varies from $-10^{\circ}$ to $10^{\circ}$. We also selected several angles ($\theta =-10^{\circ}, -5^{\circ}, 5^{\circ}, 10^{\circ}$) as examples and plotted the distribution of the corner states in real space for these cases [see Fig.~\ref{figure5}(a)], confirming that the corner modes are indeed localized at the square disk corners. These findings indicate that the material retains its SOTI state even if the ligands are rotated.

The properties of 2D materials can be controlled through external fields, such as electric field. In this work, we apply a vertical electric field to monolayer Cr(pyz)$_2$ to study its effect on the corner state. We applied an electric field ranging from -0.06 eV/\AA~ to 0.06 eV/\AA~ to the material. The plus and minus signs indicate the direction of the electric fields, which are parallel to each other but in opposite directions. By calculating the band structures after applying the electric field, it can be found that the energy positions of CBM and VBM remain largely unchanged [see Fig.~\ref{figure5}(b)]. This indicates that the band gap in Cr(pyz)$_2$ persists and remains nearly constant in size. Under the condition of maintaining the band gap, we calculated the real Chern number of the material under different electric fields, and the results all show a value of \text{``}1{''}, indicating the presence of corner states. Then, to further confirm the existence of these corner states, we calculated the energy spectrum of the material under electric fields of -0.06, -0.04, -0.02, +0.02, +0.04, +0.06 eV/\AA~, finding that the 0D corner states remain identifiable within the gap. The distribution of corner states in the square disk are also shown in the Fig.~\ref{figure5}(b). These results demonstrate that the corner states in monolayer Cr(pyz)$_2$ can exist stably under the electric field, and are not overshadowed by edge or bulk states.

\section{Conclusion}

In summary, through first-principles calculations and symmetry analysis, we proposed a 2D FiM MOF material [Cr(pyz)$_2$] with stable SOTI state. The monolayer Cr(pyz)$_2$ has FiM ground state and the easy axis is along [001] direction. The spins of all Cr atoms are aligned antiparallel to those of the pyz molecules, preserving the FiM order. In magnetic ground state, the spin-up and spin-down channels display insulating behavior with large band gaps. The real Chern number for spin-up channel is \text{``}1{''}, hosting 0D corner states in real space. Most importantly, these topological corner states are so stable that they persist even when artificial defects are created to break the symmetries. Furthermore, under other external perturbations including uniaxial/biaxial strains, ligand rotation and electric fields, the corner states also persist. Our work is the first to discover such a stable 2D FiM SOTI in MOF material, which is highly favorable for future experimental observation and holds significant potential for application in nanospintronic devices.

\section{Acknowledgments}

This work is supported by National Natural Science Foundation of China (Grant No. 12304081), the Hebei Natural Science Foundation (S$\&$T Program of Hebei, Grant No. A2023202032), the Hebei Province Major Science and Technology Support Plan Frontier Technology Project (Grant No. 242Q1401Z).

\bibliographystyle{apsrev4-1}
\bibliography{ref}

\begin{thebibliography}{46}%
\makeatletter
\providecommand \@ifxundefined [1]{%
 \@ifx{#1\undefined}
}%
\providecommand \@ifnum [1]{%
 \ifnum #1\expandafter \@firstoftwo
 \else \expandafter \@secondoftwo
 \fi
}%
\providecommand \@ifx [1]{%
 \ifx #1\expandafter \@firstoftwo
 \else \expandafter \@secondoftwo
 \fi
}%
\providecommand \natexlab [1]{#1}%
\providecommand \enquote  [1]{``#1''}%
\providecommand \bibnamefont  [1]{#1}%
\providecommand \bibfnamefont [1]{#1}%
\providecommand \citenamefont [1]{#1}%
\providecommand \href@noop [0]{\@secondoftwo}%
\providecommand \href [0]{\begingroup \@sanitize@url \@href}%
\providecommand \@href[1]{\@@startlink{#1}\@@href}%
\providecommand \@@href[1]{\endgroup#1\@@endlink}%
\providecommand \@sanitize@url [0]{\catcode `\\12\catcode `\$12\catcode
  `\&12\catcode `\#12\catcode `\^12\catcode `\_12\catcode `\%12\relax}%
\providecommand \@@startlink[1]{}%
\providecommand \@@endlink[0]{}%
\providecommand \url  [0]{\begingroup\@sanitize@url \@url }%
\providecommand \@url [1]{\endgroup\@href {#1}{\urlprefix }}%
\providecommand \urlprefix  [0]{URL }%
\providecommand \Eprint [0]{\href }%
\providecommand \doibase [0]{http://dx.doi.org/}%
\providecommand \selectlanguage [0]{\@gobble}%
\providecommand \bibinfo  [0]{\@secondoftwo}%
\providecommand \bibfield  [0]{\@secondoftwo}%
\providecommand \translation [1]{[#1]}%
\providecommand \BibitemOpen [0]{}%
\providecommand \bibitemStop [0]{}%
\providecommand \bibitemNoStop [0]{.\EOS\space}%
\providecommand \EOS [0]{\spacefactor3000\relax}%
\providecommand \BibitemShut  [1]{\csname bibitem#1\endcsname}%
\let\auto@bib@innerbib\@empty
%</preamble>
\bibitem [{\citenamefont {Hasan}\ and\ \citenamefont
  {Kane}(2010)}]{hasan2010colloquium}%
  \BibitemOpen
  \bibfield  {author} {\bibinfo {author} {\bibfnamefont {M.~Z.}\ \bibnamefont
  {Hasan}}\ and\ \bibinfo {author} {\bibfnamefont {C.~L.}\ \bibnamefont
  {Kane}},\ }\href@noop {} {\bibfield  {journal} {\bibinfo  {journal} {Rev.
  Mod. Phys.}\ }\textbf {\bibinfo {volume} {82}},\ \bibinfo {pages} {3045}
  (\bibinfo {year} {2010})}\BibitemShut {NoStop}%
\bibitem [{\citenamefont {Qi}\ and\ \citenamefont
  {Zhang}(2011)}]{qi2011topological}%
  \BibitemOpen
  \bibfield  {author} {\bibinfo {author} {\bibfnamefont {X.-L.}\ \bibnamefont
  {Qi}}\ and\ \bibinfo {author} {\bibfnamefont {S.-C.}\ \bibnamefont {Zhang}},\
  }\href@noop {} {\bibfield  {journal} {\bibinfo  {journal} {Rev. Mod. Phys.}\
  }\textbf {\bibinfo {volume} {83}},\ \bibinfo {pages} {1057} (\bibinfo {year}
  {2011})}\BibitemShut {NoStop}%
\bibitem [{\citenamefont {Zhang}\ \emph {et~al.}(2009)\citenamefont {Zhang},
  \citenamefont {Liu}, \citenamefont {Qi}, \citenamefont {Dai}, \citenamefont
  {Fang},\ and\ \citenamefont {Zhang}}]{zhang2009topological}%
  \BibitemOpen
  \bibfield  {author} {\bibinfo {author} {\bibfnamefont {H.}~\bibnamefont
  {Zhang}}, \bibinfo {author} {\bibfnamefont {C.-X.}\ \bibnamefont {Liu}},
  \bibinfo {author} {\bibfnamefont {X.-L.}\ \bibnamefont {Qi}}, \bibinfo
  {author} {\bibfnamefont {X.}~\bibnamefont {Dai}}, \bibinfo {author}
  {\bibfnamefont {Z.}~\bibnamefont {Fang}}, \ and\ \bibinfo {author}
  {\bibfnamefont {S.-C.}\ \bibnamefont {Zhang}},\ }\href@noop {} {\bibfield
  {journal} {\bibinfo  {journal} {Nat. phys.}\ }\textbf {\bibinfo {volume}
  {5}},\ \bibinfo {pages} {438} (\bibinfo {year} {2009})}\BibitemShut {NoStop}%
\bibitem [{\citenamefont {Liu}\ \emph {et~al.}(2010)\citenamefont {Liu},
  \citenamefont {Qi}, \citenamefont {Zhang}, \citenamefont {Dai}, \citenamefont
  {Fang},\ and\ \citenamefont {Zhang}}]{liu2010model}%
  \BibitemOpen
  \bibfield  {author} {\bibinfo {author} {\bibfnamefont {C.-X.}\ \bibnamefont
  {Liu}}, \bibinfo {author} {\bibfnamefont {X.-L.}\ \bibnamefont {Qi}},
  \bibinfo {author} {\bibfnamefont {H.}~\bibnamefont {Zhang}}, \bibinfo
  {author} {\bibfnamefont {X.}~\bibnamefont {Dai}}, \bibinfo {author}
  {\bibfnamefont {Z.}~\bibnamefont {Fang}}, \ and\ \bibinfo {author}
  {\bibfnamefont {S.-C.}\ \bibnamefont {Zhang}},\ }\href@noop {} {\bibfield
  {journal} {\bibinfo  {journal} {Phys. Rev. B}\ }\textbf {\bibinfo {volume}
  {82}},\ \bibinfo {pages} {045122} (\bibinfo {year} {2010})}\BibitemShut
  {NoStop}%
\bibitem [{\citenamefont {Xia}\ \emph {et~al.}(2009)\citenamefont {Xia},
  \citenamefont {Qian}, \citenamefont {Hsieh}, \citenamefont {Wray},
  \citenamefont {Pal}, \citenamefont {Lin}, \citenamefont {Bansil},
  \citenamefont {Grauer}, \citenamefont {Hor}, \citenamefont {Cava} \emph
  {et~al.}}]{xia2009observation}%
  \BibitemOpen
  \bibfield  {author} {\bibinfo {author} {\bibfnamefont {Y.}~\bibnamefont
  {Xia}}, \bibinfo {author} {\bibfnamefont {D.}~\bibnamefont {Qian}}, \bibinfo
  {author} {\bibfnamefont {D.}~\bibnamefont {Hsieh}}, \bibinfo {author}
  {\bibfnamefont {L.}~\bibnamefont {Wray}}, \bibinfo {author} {\bibfnamefont
  {A.}~\bibnamefont {Pal}}, \bibinfo {author} {\bibfnamefont {H.}~\bibnamefont
  {Lin}}, \bibinfo {author} {\bibfnamefont {A.}~\bibnamefont {Bansil}},
  \bibinfo {author} {\bibfnamefont {D.}~\bibnamefont {Grauer}}, \bibinfo
  {author} {\bibfnamefont {Y.~S.}\ \bibnamefont {Hor}}, \bibinfo {author}
  {\bibfnamefont {R.~J.}\ \bibnamefont {Cava}},  \emph {et~al.},\ }\href@noop
  {} {\bibfield  {journal} {\bibinfo  {journal} {Nat. phys.}\ }\textbf
  {\bibinfo {volume} {5}},\ \bibinfo {pages} {398} (\bibinfo {year}
  {2009})}\BibitemShut {NoStop}%
\bibitem [{\citenamefont {Hsieh}\ \emph {et~al.}(2009)\citenamefont {Hsieh},
  \citenamefont {Xia}, \citenamefont {Qian}, \citenamefont {Wray},
  \citenamefont {Dil}, \citenamefont {Meier}, \citenamefont {Osterwalder},
  \citenamefont {Patthey}, \citenamefont {Checkelsky}, \citenamefont {Ong}
  \emph {et~al.}}]{hsieh2009tunable}%
  \BibitemOpen
  \bibfield  {author} {\bibinfo {author} {\bibfnamefont {D.}~\bibnamefont
  {Hsieh}}, \bibinfo {author} {\bibfnamefont {Y.}~\bibnamefont {Xia}}, \bibinfo
  {author} {\bibfnamefont {D.}~\bibnamefont {Qian}}, \bibinfo {author}
  {\bibfnamefont {L.}~\bibnamefont {Wray}}, \bibinfo {author} {\bibfnamefont
  {J.}~\bibnamefont {Dil}}, \bibinfo {author} {\bibfnamefont {F.}~\bibnamefont
  {Meier}}, \bibinfo {author} {\bibfnamefont {J.}~\bibnamefont {Osterwalder}},
  \bibinfo {author} {\bibfnamefont {L.}~\bibnamefont {Patthey}}, \bibinfo
  {author} {\bibfnamefont {J.}~\bibnamefont {Checkelsky}}, \bibinfo {author}
  {\bibfnamefont {N.~P.}\ \bibnamefont {Ong}},  \emph {et~al.},\ }\href@noop {}
  {\bibfield  {journal} {\bibinfo  {journal} {Nature}\ }\textbf {\bibinfo
  {volume} {460}},\ \bibinfo {pages} {1101} (\bibinfo {year}
  {2009})}\BibitemShut {NoStop}%
\bibitem [{\citenamefont {Sheng}\ \emph
  {et~al.}(2019{\natexlab{a}})\citenamefont {Sheng}, \citenamefont {Chen},
  \citenamefont {Liu}, \citenamefont {Chen}, \citenamefont {Yu}, \citenamefont
  {Zhao},\ and\ \citenamefont {Yang}}]{sheng2019two1}%
  \BibitemOpen
  \bibfield  {author} {\bibinfo {author} {\bibfnamefont {X.-L.}\ \bibnamefont
  {Sheng}}, \bibinfo {author} {\bibfnamefont {C.}~\bibnamefont {Chen}},
  \bibinfo {author} {\bibfnamefont {H.}~\bibnamefont {Liu}}, \bibinfo {author}
  {\bibfnamefont {Z.}~\bibnamefont {Chen}}, \bibinfo {author} {\bibfnamefont
  {Z.-M.}\ \bibnamefont {Yu}}, \bibinfo {author} {\bibfnamefont
  {Y.}~\bibnamefont {Zhao}}, \ and\ \bibinfo {author} {\bibfnamefont {S.~A.}\
  \bibnamefont {Yang}},\ }\href@noop {} {\bibfield  {journal} {\bibinfo
  {journal} {Phys. Rev. Lett.}\ }\textbf {\bibinfo {volume} {123}},\ \bibinfo
  {pages} {256402} (\bibinfo {year} {2019}{\natexlab{a}})}\BibitemShut
  {NoStop}%
\bibitem [{\citenamefont {Chen}\ \emph {et~al.}(2021)\citenamefont {Chen},
  \citenamefont {Wu}, \citenamefont {Yu}, \citenamefont {Chen}, \citenamefont
  {Zhao}, \citenamefont {Sheng},\ and\ \citenamefont
  {Yang}}]{chen2021graphyne}%
  \BibitemOpen
  \bibfield  {author} {\bibinfo {author} {\bibfnamefont {C.}~\bibnamefont
  {Chen}}, \bibinfo {author} {\bibfnamefont {W.}~\bibnamefont {Wu}}, \bibinfo
  {author} {\bibfnamefont {Z.-M.}\ \bibnamefont {Yu}}, \bibinfo {author}
  {\bibfnamefont {Z.}~\bibnamefont {Chen}}, \bibinfo {author} {\bibfnamefont
  {Y.}~\bibnamefont {Zhao}}, \bibinfo {author} {\bibfnamefont {X.-L.}\
  \bibnamefont {Sheng}}, \ and\ \bibinfo {author} {\bibfnamefont {S.~A.}\
  \bibnamefont {Yang}},\ }\href@noop {} {\bibfield  {journal} {\bibinfo
  {journal} {Phys. Rev. B}\ }\textbf {\bibinfo {volume} {104}},\ \bibinfo
  {pages} {085205} (\bibinfo {year} {2021})}\BibitemShut {NoStop}%
\bibitem [{\citenamefont {Qian}\ \emph {et~al.}(2021)\citenamefont {Qian},
  \citenamefont {Liu},\ and\ \citenamefont {Yao}}]{qian2021second}%
  \BibitemOpen
  \bibfield  {author} {\bibinfo {author} {\bibfnamefont {S.}~\bibnamefont
  {Qian}}, \bibinfo {author} {\bibfnamefont {C.-C.}\ \bibnamefont {Liu}}, \
  and\ \bibinfo {author} {\bibfnamefont {Y.}~\bibnamefont {Yao}},\ }\href@noop
  {} {\bibfield  {journal} {\bibinfo  {journal} {Phys. Rev. B}\ }\textbf
  {\bibinfo {volume} {104}},\ \bibinfo {pages} {245427} (\bibinfo {year}
  {2021})}\BibitemShut {NoStop}%
\bibitem [{\citenamefont {Pan}\ \emph {et~al.}(2022)\citenamefont {Pan},
  \citenamefont {Li}, \citenamefont {Fan},\ and\ \citenamefont
  {Huang}}]{pan2022two}%
  \BibitemOpen
  \bibfield  {author} {\bibinfo {author} {\bibfnamefont {M.}~\bibnamefont
  {Pan}}, \bibinfo {author} {\bibfnamefont {D.}~\bibnamefont {Li}}, \bibinfo
  {author} {\bibfnamefont {J.}~\bibnamefont {Fan}}, \ and\ \bibinfo {author}
  {\bibfnamefont {H.}~\bibnamefont {Huang}},\ }\href@noop {} {\bibfield
  {journal} {\bibinfo  {journal} {npj Comput. Mater.}\ }\textbf {\bibinfo
  {volume} {8}},\ \bibinfo {pages} {1} (\bibinfo {year} {2022})}\BibitemShut
  {NoStop}%
\bibitem [{\citenamefont {Benalcazar}\ \emph {et~al.}(2019)\citenamefont
  {Benalcazar}, \citenamefont {Li},\ and\ \citenamefont
  {Hughes}}]{benalcazar2019quantization}%
  \BibitemOpen
  \bibfield  {author} {\bibinfo {author} {\bibfnamefont {W.~A.}\ \bibnamefont
  {Benalcazar}}, \bibinfo {author} {\bibfnamefont {T.}~\bibnamefont {Li}}, \
  and\ \bibinfo {author} {\bibfnamefont {T.~L.}\ \bibnamefont {Hughes}},\
  }\href@noop {} {\bibfield  {journal} {\bibinfo  {journal} {Phys. Rev. B}\
  }\textbf {\bibinfo {volume} {99}},\ \bibinfo {pages} {245151} (\bibinfo
  {year} {2019})}\BibitemShut {NoStop}%
\bibitem [{\citenamefont {Takahashi}\ \emph {et~al.}(2021)\citenamefont
  {Takahashi}, \citenamefont {Zhang},\ and\ \citenamefont
  {Murakami}}]{takahashi2021general}%
  \BibitemOpen
  \bibfield  {author} {\bibinfo {author} {\bibfnamefont {R.}~\bibnamefont
  {Takahashi}}, \bibinfo {author} {\bibfnamefont {T.}~\bibnamefont {Zhang}}, \
  and\ \bibinfo {author} {\bibfnamefont {S.}~\bibnamefont {Murakami}},\
  }\href@noop {} {\bibfield  {journal} {\bibinfo  {journal} {Phys. Rev. B}\
  }\textbf {\bibinfo {volume} {103}},\ \bibinfo {pages} {205123} (\bibinfo
  {year} {2021})}\BibitemShut {NoStop}%
\bibitem [{\citenamefont {Schindler}\ \emph {et~al.}(2018)\citenamefont
  {Schindler}, \citenamefont {Cook}, \citenamefont {Vergniory}, \citenamefont
  {Wang}, \citenamefont {Parkin}, \citenamefont {Bernevig},\ and\ \citenamefont
  {Neupert}}]{schindler2018higher}%
  \BibitemOpen
  \bibfield  {author} {\bibinfo {author} {\bibfnamefont {F.}~\bibnamefont
  {Schindler}}, \bibinfo {author} {\bibfnamefont {A.~M.}\ \bibnamefont {Cook}},
  \bibinfo {author} {\bibfnamefont {M.~G.}\ \bibnamefont {Vergniory}}, \bibinfo
  {author} {\bibfnamefont {Z.}~\bibnamefont {Wang}}, \bibinfo {author}
  {\bibfnamefont {S.~S.}\ \bibnamefont {Parkin}}, \bibinfo {author}
  {\bibfnamefont {B.~A.}\ \bibnamefont {Bernevig}}, \ and\ \bibinfo {author}
  {\bibfnamefont {T.}~\bibnamefont {Neupert}},\ }\href@noop {} {\bibfield
  {journal} {\bibinfo  {journal} {Sci. Adv.}\ }\textbf {\bibinfo {volume}
  {4}},\ \bibinfo {pages} {eaat0346} (\bibinfo {year} {2018})}\BibitemShut
  {NoStop}%
\bibitem [{\citenamefont {Zhao}\ \emph {et~al.}(2023)\citenamefont {Zhao},
  \citenamefont {Yang}, \citenamefont {Xu}, \citenamefont {Du}, \citenamefont
  {Li}, \citenamefont {Zhai}, \citenamefont {Peng}, \citenamefont {Pei},
  \citenamefont {Gao}, \citenamefont {Li} \emph
  {et~al.}}]{zhao2023topological}%
  \BibitemOpen
  \bibfield  {author} {\bibinfo {author} {\bibfnamefont {W.}~\bibnamefont
  {Zhao}}, \bibinfo {author} {\bibfnamefont {M.}~\bibnamefont {Yang}}, \bibinfo
  {author} {\bibfnamefont {R.}~\bibnamefont {Xu}}, \bibinfo {author}
  {\bibfnamefont {X.}~\bibnamefont {Du}}, \bibinfo {author} {\bibfnamefont
  {Y.}~\bibnamefont {Li}}, \bibinfo {author} {\bibfnamefont {K.}~\bibnamefont
  {Zhai}}, \bibinfo {author} {\bibfnamefont {C.}~\bibnamefont {Peng}}, \bibinfo
  {author} {\bibfnamefont {D.}~\bibnamefont {Pei}}, \bibinfo {author}
  {\bibfnamefont {H.}~\bibnamefont {Gao}}, \bibinfo {author} {\bibfnamefont
  {Y.}~\bibnamefont {Li}},  \emph {et~al.},\ }\href@noop {} {\bibfield
  {journal} {\bibinfo  {journal} {Nat. Commun.}\ }\textbf {\bibinfo {volume}
  {14}},\ \bibinfo {pages} {8089} (\bibinfo {year} {2023})}\BibitemShut
  {NoStop}%
\bibitem [{\citenamefont {Sheng}\ \emph
  {et~al.}(2019{\natexlab{b}})\citenamefont {Sheng}, \citenamefont {Chen},
  \citenamefont {Liu}, \citenamefont {Chen}, \citenamefont {Yu}, \citenamefont
  {Zhao},\ and\ \citenamefont {Yang}}]{sheng2019two2}%
  \BibitemOpen
  \bibfield  {author} {\bibinfo {author} {\bibfnamefont {X.-L.}\ \bibnamefont
  {Sheng}}, \bibinfo {author} {\bibfnamefont {C.}~\bibnamefont {Chen}},
  \bibinfo {author} {\bibfnamefont {H.}~\bibnamefont {Liu}}, \bibinfo {author}
  {\bibfnamefont {Z.}~\bibnamefont {Chen}}, \bibinfo {author} {\bibfnamefont
  {Z.-M.}\ \bibnamefont {Yu}}, \bibinfo {author} {\bibfnamefont
  {Y.}~\bibnamefont {Zhao}}, \ and\ \bibinfo {author} {\bibfnamefont {S.~A.}\
  \bibnamefont {Yang}},\ }\href@noop {} {\bibfield  {journal} {\bibinfo
  {journal} {Phys. Rev. Lett.}\ }\textbf {\bibinfo {volume} {123}},\ \bibinfo
  {pages} {256402} (\bibinfo {year} {2019}{\natexlab{b}})}\BibitemShut
  {NoStop}%
\bibitem [{\citenamefont {Costa}\ \emph {et~al.}(2021)\citenamefont {Costa},
  \citenamefont {Schleder}, \citenamefont {Mera~Acosta}, \citenamefont
  {Padilha}, \citenamefont {Cerasoli}, \citenamefont {Buongiorno~Nardelli},\
  and\ \citenamefont {Fazzio}}]{costa2021discovery}%
  \BibitemOpen
  \bibfield  {author} {\bibinfo {author} {\bibfnamefont {M.}~\bibnamefont
  {Costa}}, \bibinfo {author} {\bibfnamefont {G.~R.}\ \bibnamefont {Schleder}},
  \bibinfo {author} {\bibfnamefont {C.}~\bibnamefont {Mera~Acosta}}, \bibinfo
  {author} {\bibfnamefont {A.~C.}\ \bibnamefont {Padilha}}, \bibinfo {author}
  {\bibfnamefont {F.}~\bibnamefont {Cerasoli}}, \bibinfo {author}
  {\bibfnamefont {M.}~\bibnamefont {Buongiorno~Nardelli}}, \ and\ \bibinfo
  {author} {\bibfnamefont {A.}~\bibnamefont {Fazzio}},\ }\href@noop {}
  {\bibfield  {journal} {\bibinfo  {journal} {npj Comput. Mater.}\ }\textbf
  {\bibinfo {volume} {7}},\ \bibinfo {pages} {49} (\bibinfo {year}
  {2021})}\BibitemShut {NoStop}%
\bibitem [{\citenamefont {He}\ \emph {et~al.}(2023)\citenamefont {He},
  \citenamefont {Zhang}, \citenamefont {Li}, \citenamefont {Jin}, \citenamefont
  {Liu}, \citenamefont {Liu},\ and\ \citenamefont {Yuan}}]{he2023metal}%
  \BibitemOpen
  \bibfield  {author} {\bibinfo {author} {\bibfnamefont {T.}~\bibnamefont
  {He}}, \bibinfo {author} {\bibfnamefont {X.}~\bibnamefont {Zhang}}, \bibinfo
  {author} {\bibfnamefont {Y.}~\bibnamefont {Li}}, \bibinfo {author}
  {\bibfnamefont {L.}~\bibnamefont {Jin}}, \bibinfo {author} {\bibfnamefont
  {Y.}~\bibnamefont {Liu}}, \bibinfo {author} {\bibfnamefont {G.}~\bibnamefont
  {Liu}}, \ and\ \bibinfo {author} {\bibfnamefont {H.}~\bibnamefont {Yuan}},\
  }\href@noop {} {\bibfield  {journal} {\bibinfo  {journal} {Mater. Today
  Nano}\ }\textbf {\bibinfo {volume} {24}},\ \bibinfo {pages} {100389}
  (\bibinfo {year} {2023})}\BibitemShut {NoStop}%
\bibitem [{\citenamefont {Ezawa}(2018)}]{ezawa2018topological}%
  \BibitemOpen
  \bibfield  {author} {\bibinfo {author} {\bibfnamefont {M.}~\bibnamefont
  {Ezawa}},\ }\href@noop {} {\bibfield  {journal} {\bibinfo  {journal} {Phys.
  Rev. Lett.}\ }\textbf {\bibinfo {volume} {121}},\ \bibinfo {pages} {116801}
  (\bibinfo {year} {2018})}\BibitemShut {NoStop}%
\bibitem [{\citenamefont {Yan}\ \emph {et~al.}(2022)\citenamefont {Yan},
  \citenamefont {Tian}, \citenamefont {Yan},\ and\ \citenamefont
  {Chen}}]{yan2022plane}%
  \BibitemOpen
  \bibfield  {author} {\bibinfo {author} {\bibfnamefont {X.}~\bibnamefont
  {Yan}}, \bibinfo {author} {\bibfnamefont {M.}~\bibnamefont {Tian}}, \bibinfo
  {author} {\bibfnamefont {J.}~\bibnamefont {Yan}}, \ and\ \bibinfo {author}
  {\bibfnamefont {Q.}~\bibnamefont {Chen}},\ }\href@noop {} {\bibfield
  {journal} {\bibinfo  {journal} {Physica B}\ }\textbf {\bibinfo {volume}
  {630}},\ \bibinfo {pages} {413641} (\bibinfo {year} {2022})}\BibitemShut
  {NoStop}%
\bibitem [{\citenamefont {Guo}\ \emph {et~al.}(2023)\citenamefont {Guo},
  \citenamefont {Liu}, \citenamefont {Jiang}, \citenamefont {Zhang},
  \citenamefont {Jin}, \citenamefont {Liu},\ and\ \citenamefont
  {Liu}}]{guo2023magnetic}%
  \BibitemOpen
  \bibfield  {author} {\bibinfo {author} {\bibfnamefont {Z.}~\bibnamefont
  {Guo}}, \bibinfo {author} {\bibfnamefont {Y.}~\bibnamefont {Liu}}, \bibinfo
  {author} {\bibfnamefont {H.}~\bibnamefont {Jiang}}, \bibinfo {author}
  {\bibfnamefont {X.}~\bibnamefont {Zhang}}, \bibinfo {author} {\bibfnamefont
  {L.}~\bibnamefont {Jin}}, \bibinfo {author} {\bibfnamefont {C.}~\bibnamefont
  {Liu}}, \ and\ \bibinfo {author} {\bibfnamefont {G.}~\bibnamefont {Liu}},\
  }\href@noop {} {\bibfield  {journal} {\bibinfo  {journal} {Mater. Today
  Phys.}\ }\textbf {\bibinfo {volume} {36}},\ \bibinfo {pages} {101153}
  (\bibinfo {year} {2023})}\BibitemShut {NoStop}%
\bibitem [{\citenamefont {Xu}\ \emph {et~al.}(2019)\citenamefont {Xu},
  \citenamefont {Song}, \citenamefont {Wang}, \citenamefont {Weng},\ and\
  \citenamefont {Dai}}]{xu2019higher}%
  \BibitemOpen
  \bibfield  {author} {\bibinfo {author} {\bibfnamefont {Y.}~\bibnamefont
  {Xu}}, \bibinfo {author} {\bibfnamefont {Z.}~\bibnamefont {Song}}, \bibinfo
  {author} {\bibfnamefont {Z.}~\bibnamefont {Wang}}, \bibinfo {author}
  {\bibfnamefont {H.}~\bibnamefont {Weng}}, \ and\ \bibinfo {author}
  {\bibfnamefont {X.}~\bibnamefont {Dai}},\ }\href@noop {} {\bibfield
  {journal} {\bibinfo  {journal} {Phys. Rev. Lett.}\ }\textbf {\bibinfo
  {volume} {122}},\ \bibinfo {pages} {256402} (\bibinfo {year}
  {2019})}\BibitemShut {NoStop}%
\bibitem [{\citenamefont {Chen}\ \emph {et~al.}(2020)\citenamefont {Chen},
  \citenamefont {Song}, \citenamefont {Zhao}, \citenamefont {Chen},
  \citenamefont {Yu}, \citenamefont {Sheng},\ and\ \citenamefont
  {Yang}}]{chen2020universal}%
  \BibitemOpen
  \bibfield  {author} {\bibinfo {author} {\bibfnamefont {C.}~\bibnamefont
  {Chen}}, \bibinfo {author} {\bibfnamefont {Z.}~\bibnamefont {Song}}, \bibinfo
  {author} {\bibfnamefont {J.-Z.}\ \bibnamefont {Zhao}}, \bibinfo {author}
  {\bibfnamefont {Z.}~\bibnamefont {Chen}}, \bibinfo {author} {\bibfnamefont
  {Z.-M.}\ \bibnamefont {Yu}}, \bibinfo {author} {\bibfnamefont {X.-L.}\
  \bibnamefont {Sheng}}, \ and\ \bibinfo {author} {\bibfnamefont {S.~A.}\
  \bibnamefont {Yang}},\ }\href@noop {} {\bibfield  {journal} {\bibinfo
  {journal} {Phys. Rev. Lett.}\ }\textbf {\bibinfo {volume} {125}},\ \bibinfo
  {pages} {056402} (\bibinfo {year} {2020})}\BibitemShut {NoStop}%
\bibitem [{\citenamefont {Mao}\ \emph {et~al.}(2021)\citenamefont {Mao},
  \citenamefont {Hu}, \citenamefont {Wang}, \citenamefont {Dai}, \citenamefont
  {Huang}, \citenamefont {Mokrousov},\ and\ \citenamefont
  {Niu}}]{mao2021magnetism}%
  \BibitemOpen
  \bibfield  {author} {\bibinfo {author} {\bibfnamefont {N.}~\bibnamefont
  {Mao}}, \bibinfo {author} {\bibfnamefont {X.}~\bibnamefont {Hu}}, \bibinfo
  {author} {\bibfnamefont {H.}~\bibnamefont {Wang}}, \bibinfo {author}
  {\bibfnamefont {Y.}~\bibnamefont {Dai}}, \bibinfo {author} {\bibfnamefont
  {B.}~\bibnamefont {Huang}}, \bibinfo {author} {\bibfnamefont
  {Y.}~\bibnamefont {Mokrousov}}, \ and\ \bibinfo {author} {\bibfnamefont
  {C.}~\bibnamefont {Niu}},\ }\href@noop {} {\bibfield  {journal} {\bibinfo
  {journal} {Phys. Rev. B}\ }\textbf {\bibinfo {volume} {103}},\ \bibinfo
  {pages} {195152} (\bibinfo {year} {2021})}\BibitemShut {NoStop}%
\bibitem [{\citenamefont {Sheng}\ \emph {et~al.}(2022)\citenamefont {Sheng},
  \citenamefont {Zhang}, \citenamefont {Yuan},\ and\ \citenamefont
  {Wang}}]{sheng2022strain}%
  \BibitemOpen
  \bibfield  {author} {\bibinfo {author} {\bibfnamefont {K.}~\bibnamefont
  {Sheng}}, \bibinfo {author} {\bibfnamefont {B.}~\bibnamefont {Zhang}},
  \bibinfo {author} {\bibfnamefont {H.-K.}\ \bibnamefont {Yuan}}, \ and\
  \bibinfo {author} {\bibfnamefont {Z.-Y.}\ \bibnamefont {Wang}},\ }\href@noop
  {} {\bibfield  {journal} {\bibinfo  {journal} {Phys. Rev. B}\ }\textbf
  {\bibinfo {volume} {105}},\ \bibinfo {pages} {195312} (\bibinfo {year}
  {2022})}\BibitemShut {NoStop}%
\bibitem [{\citenamefont {Wang}\ \emph {et~al.}(2023)\citenamefont {Wang},
  \citenamefont {Li}, \citenamefont {Li}, \citenamefont {Xie}, \citenamefont
  {Wang}, \citenamefont {Yuan}, \citenamefont {Wang}, \citenamefont {Cheng},
  \citenamefont {Yu},\ and\ \citenamefont {Zhang}}]{wang2023magnetic}%
  \BibitemOpen
  \bibfield  {author} {\bibinfo {author} {\bibfnamefont {X.}~\bibnamefont
  {Wang}}, \bibinfo {author} {\bibfnamefont {X.-P.}\ \bibnamefont {Li}},
  \bibinfo {author} {\bibfnamefont {J.}~\bibnamefont {Li}}, \bibinfo {author}
  {\bibfnamefont {C.}~\bibnamefont {Xie}}, \bibinfo {author} {\bibfnamefont
  {J.}~\bibnamefont {Wang}}, \bibinfo {author} {\bibfnamefont {H.}~\bibnamefont
  {Yuan}}, \bibinfo {author} {\bibfnamefont {W.}~\bibnamefont {Wang}}, \bibinfo
  {author} {\bibfnamefont {Z.}~\bibnamefont {Cheng}}, \bibinfo {author}
  {\bibfnamefont {Z.-M.}\ \bibnamefont {Yu}}, \ and\ \bibinfo {author}
  {\bibfnamefont {G.}~\bibnamefont {Zhang}},\ }\href@noop {} {\bibfield
  {journal} {\bibinfo  {journal} {Adv. Funct. Mater.}\ }\textbf {\bibinfo
  {volume} {33}},\ \bibinfo {pages} {2304499} (\bibinfo {year}
  {2023})}\BibitemShut {NoStop}%
\bibitem [{\citenamefont {Mu}\ \emph {et~al.}(2022)\citenamefont {Mu},
  \citenamefont {Zhao}, \citenamefont {Zhang},\ and\ \citenamefont
  {Wang}}]{mu2022antiferromagnetic}%
  \BibitemOpen
  \bibfield  {author} {\bibinfo {author} {\bibfnamefont {H.}~\bibnamefont
  {Mu}}, \bibinfo {author} {\bibfnamefont {G.}~\bibnamefont {Zhao}}, \bibinfo
  {author} {\bibfnamefont {H.}~\bibnamefont {Zhang}}, \ and\ \bibinfo {author}
  {\bibfnamefont {Z.}~\bibnamefont {Wang}},\ }\href@noop {} {\bibfield
  {journal} {\bibinfo  {journal} {npj Comput. Mater.}\ }\textbf {\bibinfo
  {volume} {8}},\ \bibinfo {pages} {82} (\bibinfo {year} {2022})}\BibitemShut
  {NoStop}%
\bibitem [{\citenamefont {Liu}\ \emph {et~al.}(2023)\citenamefont {Liu},
  \citenamefont {Jiang}, \citenamefont {Guo}, \citenamefont {Zhang},
  \citenamefont {Jin}, \citenamefont {Liu},\ and\ \citenamefont
  {Liu}}]{liu2023magnetic}%
  \BibitemOpen
  \bibfield  {author} {\bibinfo {author} {\bibfnamefont {G.}~\bibnamefont
  {Liu}}, \bibinfo {author} {\bibfnamefont {H.}~\bibnamefont {Jiang}}, \bibinfo
  {author} {\bibfnamefont {Z.}~\bibnamefont {Guo}}, \bibinfo {author}
  {\bibfnamefont {X.}~\bibnamefont {Zhang}}, \bibinfo {author} {\bibfnamefont
  {L.}~\bibnamefont {Jin}}, \bibinfo {author} {\bibfnamefont {C.}~\bibnamefont
  {Liu}}, \ and\ \bibinfo {author} {\bibfnamefont {Y.}~\bibnamefont {Liu}},\
  }\href@noop {} {\bibfield  {journal} {\bibinfo  {journal} {Adv. Sci.}\
  }\textbf {\bibinfo {volume} {10}},\ \bibinfo {pages} {2301952} (\bibinfo
  {year} {2023})}\BibitemShut {NoStop}%
\bibitem [{\citenamefont {Zhang}\ \emph
  {et~al.}(2023{\natexlab{a}})\citenamefont {Zhang}, \citenamefont {He},
  \citenamefont {Liu}, \citenamefont {Dai}, \citenamefont {Liu}, \citenamefont
  {Chen}, \citenamefont {Wu}, \citenamefont {Zhu},\ and\ \citenamefont
  {Yang}}]{zhang2023magnetic}%
  \BibitemOpen
  \bibfield  {author} {\bibinfo {author} {\bibfnamefont {X.}~\bibnamefont
  {Zhang}}, \bibinfo {author} {\bibfnamefont {T.}~\bibnamefont {He}}, \bibinfo
  {author} {\bibfnamefont {Y.}~\bibnamefont {Liu}}, \bibinfo {author}
  {\bibfnamefont {X.}~\bibnamefont {Dai}}, \bibinfo {author} {\bibfnamefont
  {G.}~\bibnamefont {Liu}}, \bibinfo {author} {\bibfnamefont {C.}~\bibnamefont
  {Chen}}, \bibinfo {author} {\bibfnamefont {W.}~\bibnamefont {Wu}}, \bibinfo
  {author} {\bibfnamefont {J.}~\bibnamefont {Zhu}}, \ and\ \bibinfo {author}
  {\bibfnamefont {S.~A.}\ \bibnamefont {Yang}},\ }\href@noop {} {\bibfield
  {journal} {\bibinfo  {journal} {Nano Lett.}\ }\textbf {\bibinfo {volume}
  {23}},\ \bibinfo {pages} {7358} (\bibinfo {year}
  {2023}{\natexlab{a}})}\BibitemShut {NoStop}%
\bibitem [{\citenamefont {Dong}\ \emph {et~al.}(2021)\citenamefont {Dong},
  \citenamefont {Xu}, \citenamefont {Zhou}, \citenamefont {Cai}, \citenamefont
  {Wu}, \citenamefont {Tang},\ and\ \citenamefont
  {Jiang}}]{dong2021electrically}%
  \BibitemOpen
  \bibfield  {author} {\bibinfo {author} {\bibfnamefont {Y.}~\bibnamefont
  {Dong}}, \bibinfo {author} {\bibfnamefont {T.}~\bibnamefont {Xu}}, \bibinfo
  {author} {\bibfnamefont {H.-A.}\ \bibnamefont {Zhou}}, \bibinfo {author}
  {\bibfnamefont {L.}~\bibnamefont {Cai}}, \bibinfo {author} {\bibfnamefont
  {H.}~\bibnamefont {Wu}}, \bibinfo {author} {\bibfnamefont {J.}~\bibnamefont
  {Tang}}, \ and\ \bibinfo {author} {\bibfnamefont {W.}~\bibnamefont {Jiang}},\
  }\href@noop {} {\bibfield  {journal} {\bibinfo  {journal} {Adv. Funct.
  Mater.}\ }\textbf {\bibinfo {volume} {31}},\ \bibinfo {pages} {2007485}
  (\bibinfo {year} {2021})}\BibitemShut {NoStop}%
\bibitem [{\citenamefont {Wang}\ \emph {et~al.}(2021)\citenamefont {Wang},
  \citenamefont {Zhang}, \citenamefont {Zhang}, \citenamefont {Zheng},
  \citenamefont {Zhang}, \citenamefont {Wang}, \citenamefont {Klein},
  \citenamefont {Ravelosona},\ and\ \citenamefont {Zhao}}]{wang2021ultrafast}%
  \BibitemOpen
  \bibfield  {author} {\bibinfo {author} {\bibfnamefont {G.}~\bibnamefont
  {Wang}}, \bibinfo {author} {\bibfnamefont {Y.}~\bibnamefont {Zhang}},
  \bibinfo {author} {\bibfnamefont {Z.}~\bibnamefont {Zhang}}, \bibinfo
  {author} {\bibfnamefont {Z.}~\bibnamefont {Zheng}}, \bibinfo {author}
  {\bibfnamefont {K.}~\bibnamefont {Zhang}}, \bibinfo {author} {\bibfnamefont
  {J.}~\bibnamefont {Wang}}, \bibinfo {author} {\bibfnamefont {J.-O.}\
  \bibnamefont {Klein}}, \bibinfo {author} {\bibfnamefont {D.}~\bibnamefont
  {Ravelosona}}, \ and\ \bibinfo {author} {\bibfnamefont {W.}~\bibnamefont
  {Zhao}},\ }\href@noop {} {\bibfield  {journal} {\bibinfo  {journal} {IEEE
  Electr. Device L.}\ }\textbf {\bibinfo {volume} {42}},\ \bibinfo {pages}
  {621} (\bibinfo {year} {2021})}\BibitemShut {NoStop}%
\bibitem [{\citenamefont {Zhang}\ \emph {et~al.}(2019)\citenamefont {Zhang},
  \citenamefont {Zhu}, \citenamefont {Zhang}, \citenamefont {Zhang},
  \citenamefont {Nan}, \citenamefont {Zheng}, \citenamefont {Zhang},\ and\
  \citenamefont {Zhao}}]{zhang2019skyrmion}%
  \BibitemOpen
  \bibfield  {author} {\bibinfo {author} {\bibfnamefont {Z.}~\bibnamefont
  {Zhang}}, \bibinfo {author} {\bibfnamefont {Y.}~\bibnamefont {Zhu}}, \bibinfo
  {author} {\bibfnamefont {Y.}~\bibnamefont {Zhang}}, \bibinfo {author}
  {\bibfnamefont {K.}~\bibnamefont {Zhang}}, \bibinfo {author} {\bibfnamefont
  {J.}~\bibnamefont {Nan}}, \bibinfo {author} {\bibfnamefont {Z.}~\bibnamefont
  {Zheng}}, \bibinfo {author} {\bibfnamefont {Y.}~\bibnamefont {Zhang}}, \ and\
  \bibinfo {author} {\bibfnamefont {W.}~\bibnamefont {Zhao}},\ }\href@noop {}
  {\bibfield  {journal} {\bibinfo  {journal} {IEEE Electr. Device L.}\ }\textbf
  {\bibinfo {volume} {40}},\ \bibinfo {pages} {1984} (\bibinfo {year}
  {2019})}\BibitemShut {NoStop}%
\bibitem [{\citenamefont {Reza}\ and\ \citenamefont
  {Roy}(2019)}]{reza2019fast}%
  \BibitemOpen
  \bibfield  {author} {\bibinfo {author} {\bibfnamefont {A.~K.}\ \bibnamefont
  {Reza}}\ and\ \bibinfo {author} {\bibfnamefont {K.}~\bibnamefont {Roy}},\
  }\href@noop {} {\bibfield  {journal} {\bibinfo  {journal} {J. Appl. Phys.}\
  }\textbf {\bibinfo {volume} {126}} (\bibinfo {year} {2019})}\BibitemShut
  {NoStop}%
\bibitem [{\citenamefont {Zhang}\ \emph
  {et~al.}(2023{\natexlab{b}})\citenamefont {Zhang}, \citenamefont {Feng},
  \citenamefont {Zheng}, \citenamefont {Zhang}, \citenamefont {Lin},
  \citenamefont {Sun}, \citenamefont {Wang}, \citenamefont {Wang},
  \citenamefont {Wei}, \citenamefont {Vallobra} \emph
  {et~al.}}]{zhang2023ferrimagnets}%
  \BibitemOpen
  \bibfield  {author} {\bibinfo {author} {\bibfnamefont {Y.}~\bibnamefont
  {Zhang}}, \bibinfo {author} {\bibfnamefont {X.}~\bibnamefont {Feng}},
  \bibinfo {author} {\bibfnamefont {Z.}~\bibnamefont {Zheng}}, \bibinfo
  {author} {\bibfnamefont {Z.}~\bibnamefont {Zhang}}, \bibinfo {author}
  {\bibfnamefont {K.}~\bibnamefont {Lin}}, \bibinfo {author} {\bibfnamefont
  {X.}~\bibnamefont {Sun}}, \bibinfo {author} {\bibfnamefont {G.}~\bibnamefont
  {Wang}}, \bibinfo {author} {\bibfnamefont {J.}~\bibnamefont {Wang}}, \bibinfo
  {author} {\bibfnamefont {J.}~\bibnamefont {Wei}}, \bibinfo {author}
  {\bibfnamefont {P.}~\bibnamefont {Vallobra}},  \emph {et~al.},\ }\href@noop
  {} {\bibfield  {journal} {\bibinfo  {journal} {Appl. Phys. Rev.}\ }\textbf
  {\bibinfo {volume} {10}} (\bibinfo {year} {2023}{\natexlab{b}})}\BibitemShut
  {NoStop}%
\bibitem [{\citenamefont {Kresse}\ and\ \citenamefont
  {Furthm{\"u}ller}(1996{\natexlab{a}})}]{kresse1996efficiency1}%
  \BibitemOpen
  \bibfield  {author} {\bibinfo {author} {\bibfnamefont {G.}~\bibnamefont
  {Kresse}}\ and\ \bibinfo {author} {\bibfnamefont {J.}~\bibnamefont
  {Furthm{\"u}ller}},\ }\href@noop {} {\bibfield  {journal} {\bibinfo
  {journal} {Comput. Mater. Sci.}\ }\textbf {\bibinfo {volume} {6}},\ \bibinfo
  {pages} {15} (\bibinfo {year} {1996}{\natexlab{a}})}\BibitemShut {NoStop}%
\bibitem [{\citenamefont {Kresse}\ and\ \citenamefont
  {Furthm{\"u}ller}(1996{\natexlab{b}})}]{kresse1996efficient2}%
  \BibitemOpen
  \bibfield  {author} {\bibinfo {author} {\bibfnamefont {G.}~\bibnamefont
  {Kresse}}\ and\ \bibinfo {author} {\bibfnamefont {J.}~\bibnamefont
  {Furthm{\"u}ller}},\ }\href@noop {} {\bibfield  {journal} {\bibinfo
  {journal} {Phys. Rev. B}\ }\textbf {\bibinfo {volume} {54}},\ \bibinfo
  {pages} {11169} (\bibinfo {year} {1996}{\natexlab{b}})}\BibitemShut {NoStop}%
\bibitem [{\citenamefont {Perdew}\ \emph {et~al.}(1996)\citenamefont {Perdew},
  \citenamefont {Burke},\ and\ \citenamefont
  {Ernzerhof}}]{perdew1996generalized}%
  \BibitemOpen
  \bibfield  {author} {\bibinfo {author} {\bibfnamefont {J.~P.}\ \bibnamefont
  {Perdew}}, \bibinfo {author} {\bibfnamefont {K.}~\bibnamefont {Burke}}, \
  and\ \bibinfo {author} {\bibfnamefont {M.}~\bibnamefont {Ernzerhof}},\
  }\href@noop {} {\bibfield  {journal} {\bibinfo  {journal} {Phys. Rev. lett.}\
  }\textbf {\bibinfo {volume} {77}},\ \bibinfo {pages} {3865} (\bibinfo {year}
  {1996})}\BibitemShut {NoStop}%
\bibitem [{\citenamefont {Monkhorst}\ and\ \citenamefont
  {Pack}(1976)}]{monkhorst1976special}%
  \BibitemOpen
  \bibfield  {author} {\bibinfo {author} {\bibfnamefont {H.~J.}\ \bibnamefont
  {Monkhorst}}\ and\ \bibinfo {author} {\bibfnamefont {J.~D.}\ \bibnamefont
  {Pack}},\ }\href@noop {} {\bibfield  {journal} {\bibinfo  {journal} {Phys.
  Rev. B}\ }\textbf {\bibinfo {volume} {13}},\ \bibinfo {pages} {5188}
  (\bibinfo {year} {1976})}\BibitemShut {NoStop}%
\bibitem [{\citenamefont {Yang}\ \emph {et~al.}(1999)\citenamefont {Yang},
  \citenamefont {Huang}, \citenamefont {Ye},\ and\ \citenamefont
  {Xie}}]{yang1999influence}%
  \BibitemOpen
  \bibfield  {author} {\bibinfo {author} {\bibfnamefont {Z.}~\bibnamefont
  {Yang}}, \bibinfo {author} {\bibfnamefont {Z.}~\bibnamefont {Huang}},
  \bibinfo {author} {\bibfnamefont {L.}~\bibnamefont {Ye}}, \ and\ \bibinfo
  {author} {\bibfnamefont {X.}~\bibnamefont {Xie}},\ }\href@noop {} {\bibfield
  {journal} {\bibinfo  {journal} {Phys. Rev. B}\ }\textbf {\bibinfo {volume}
  {60}},\ \bibinfo {pages} {15674} (\bibinfo {year} {1999})}\BibitemShut
  {NoStop}%
\bibitem [{\citenamefont {Anisimov}\ \emph {et~al.}(1991)\citenamefont
  {Anisimov}, \citenamefont {Zaanen},\ and\ \citenamefont
  {Andersen}}]{anisimov1991band}%
  \BibitemOpen
  \bibfield  {author} {\bibinfo {author} {\bibfnamefont {V.~I.}\ \bibnamefont
  {Anisimov}}, \bibinfo {author} {\bibfnamefont {J.}~\bibnamefont {Zaanen}}, \
  and\ \bibinfo {author} {\bibfnamefont {O.~K.}\ \bibnamefont {Andersen}},\
  }\href@noop {} {\bibfield  {journal} {\bibinfo  {journal} {Phys. Rev. B}\
  }\textbf {\bibinfo {volume} {44}},\ \bibinfo {pages} {943} (\bibinfo {year}
  {1991})}\BibitemShut {NoStop}%
\bibitem [{\citenamefont {Li}\ \emph {et~al.}(2021)\citenamefont {Li},
  \citenamefont {Lv}, \citenamefont {Liu}, \citenamefont {Jin}, \citenamefont
  {Wu}, \citenamefont {Li},\ and\ \citenamefont {Yang}}]{li2021two}%
  \BibitemOpen
  \bibfield  {author} {\bibinfo {author} {\bibfnamefont {X.}~\bibnamefont
  {Li}}, \bibinfo {author} {\bibfnamefont {H.}~\bibnamefont {Lv}}, \bibinfo
  {author} {\bibfnamefont {X.}~\bibnamefont {Liu}}, \bibinfo {author}
  {\bibfnamefont {T.}~\bibnamefont {Jin}}, \bibinfo {author} {\bibfnamefont
  {X.}~\bibnamefont {Wu}}, \bibinfo {author} {\bibfnamefont {X.}~\bibnamefont
  {Li}}, \ and\ \bibinfo {author} {\bibfnamefont {J.}~\bibnamefont {Yang}},\
  }\href@noop {} {\bibfield  {journal} {\bibinfo  {journal} {Sci. China Chem.}\
  }\textbf {\bibinfo {volume} {64}},\ \bibinfo {pages} {2212} (\bibinfo {year}
  {2021})}\BibitemShut {NoStop}%
\bibitem [{\citenamefont {Gao}\ \emph {et~al.}(2021)\citenamefont {Gao},
  \citenamefont {Wu}, \citenamefont {Persson},\ and\ \citenamefont
  {Wang}}]{gao2021irvsp}%
  \BibitemOpen
  \bibfield  {author} {\bibinfo {author} {\bibfnamefont {J.}~\bibnamefont
  {Gao}}, \bibinfo {author} {\bibfnamefont {Q.}~\bibnamefont {Wu}}, \bibinfo
  {author} {\bibfnamefont {C.}~\bibnamefont {Persson}}, \ and\ \bibinfo
  {author} {\bibfnamefont {Z.}~\bibnamefont {Wang}},\ }\href@noop {} {\bibfield
   {journal} {\bibinfo  {journal} {Comput. Phys. Commun.}\ }\textbf {\bibinfo
  {volume} {261}},\ \bibinfo {pages} {107760} (\bibinfo {year}
  {2021})}\BibitemShut {NoStop}%
\bibitem [{\citenamefont {Pizzi}\ \emph {et~al.}(2020)\citenamefont {Pizzi},
  \citenamefont {Vitale}, \citenamefont {Arita}, \citenamefont {Bl{\"u}gel},
  \citenamefont {Freimuth}, \citenamefont {G{\'e}ranton}, \citenamefont
  {Gibertini}, \citenamefont {Gresch}, \citenamefont {Johnson}, \citenamefont
  {Koretsune} \emph {et~al.}}]{pizzi2020wannier90}%
  \BibitemOpen
  \bibfield  {author} {\bibinfo {author} {\bibfnamefont {G.}~\bibnamefont
  {Pizzi}}, \bibinfo {author} {\bibfnamefont {V.}~\bibnamefont {Vitale}},
  \bibinfo {author} {\bibfnamefont {R.}~\bibnamefont {Arita}}, \bibinfo
  {author} {\bibfnamefont {S.}~\bibnamefont {Bl{\"u}gel}}, \bibinfo {author}
  {\bibfnamefont {F.}~\bibnamefont {Freimuth}}, \bibinfo {author}
  {\bibfnamefont {G.}~\bibnamefont {G{\'e}ranton}}, \bibinfo {author}
  {\bibfnamefont {M.}~\bibnamefont {Gibertini}}, \bibinfo {author}
  {\bibfnamefont {D.}~\bibnamefont {Gresch}}, \bibinfo {author} {\bibfnamefont
  {C.}~\bibnamefont {Johnson}}, \bibinfo {author} {\bibfnamefont
  {T.}~\bibnamefont {Koretsune}},  \emph {et~al.},\ }\href@noop {} {\bibfield
  {journal} {\bibinfo  {journal} {J. Phys-Condens. Mat.}\ }\textbf {\bibinfo
  {volume} {32}},\ \bibinfo {pages} {165902} (\bibinfo {year}
  {2020})}\BibitemShut {NoStop}%
\bibitem [{\citenamefont {Moldovan}\ and\ \citenamefont
  {Anelkovi{\'c}}(2017)}]{moldovan2017peeters}%
  \BibitemOpen
  \bibfield  {author} {\bibinfo {author} {\bibfnamefont {D.}~\bibnamefont
  {Moldovan}}\ and\ \bibinfo {author} {\bibfnamefont {M.}~\bibnamefont
  {Anelkovi{\'c}}},\ }\href@noop {} {\enquote {\bibinfo {title} {Peeters, f.
  pybinding v0. 9.4: a python package for tight-binding calculations. this work
  was supported by the flemish science foundation (fwo-vl) and the methusalem
  funding of the flemish government},}\ } (\bibinfo {year} {2017})\BibitemShut
  {NoStop}%
\bibitem [{\citenamefont {Yang}\ \emph {et~al.}(2022)\citenamefont {Yang},
  \citenamefont {Ji}, \citenamefont {Feng}, \citenamefont {Chen}, \citenamefont
  {Bellaiche},\ and\ \citenamefont {Xiang}}]{yang2022two}%
  \BibitemOpen
  \bibfield  {author} {\bibinfo {author} {\bibfnamefont {Y.}~\bibnamefont
  {Yang}}, \bibinfo {author} {\bibfnamefont {J.}~\bibnamefont {Ji}}, \bibinfo
  {author} {\bibfnamefont {J.}~\bibnamefont {Feng}}, \bibinfo {author}
  {\bibfnamefont {S.}~\bibnamefont {Chen}}, \bibinfo {author} {\bibfnamefont
  {L.}~\bibnamefont {Bellaiche}}, \ and\ \bibinfo {author} {\bibfnamefont
  {H.}~\bibnamefont {Xiang}},\ }\href@noop {} {\bibfield  {journal} {\bibinfo
  {journal} {J. Am. Chem. SOC.}\ }\textbf {\bibinfo {volume} {144}},\ \bibinfo
  {pages} {14907} (\bibinfo {year} {2022})}\BibitemShut {NoStop}%
\bibitem [{\citenamefont {Lv}\ \emph {et~al.}(2022)\citenamefont {Lv},
  \citenamefont {Li}, \citenamefont {Wu}, \citenamefont {Liu}, \citenamefont
  {Li}, \citenamefont {Wu},\ and\ \citenamefont {Yang}}]{lv2022enhanced}%
  \BibitemOpen
  \bibfield  {author} {\bibinfo {author} {\bibfnamefont {H.}~\bibnamefont
  {Lv}}, \bibinfo {author} {\bibfnamefont {X.}~\bibnamefont {Li}}, \bibinfo
  {author} {\bibfnamefont {D.}~\bibnamefont {Wu}}, \bibinfo {author}
  {\bibfnamefont {Y.}~\bibnamefont {Liu}}, \bibinfo {author} {\bibfnamefont
  {X.}~\bibnamefont {Li}}, \bibinfo {author} {\bibfnamefont {X.}~\bibnamefont
  {Wu}}, \ and\ \bibinfo {author} {\bibfnamefont {J.}~\bibnamefont {Yang}},\
  }\href@noop {} {\bibfield  {journal} {\bibinfo  {journal} {Nano Lett.}\
  }\textbf {\bibinfo {volume} {22}},\ \bibinfo {pages} {1573} (\bibinfo {year}
  {2022})}\BibitemShut {NoStop}%
\bibitem [{\citenamefont {Perlepe}\ \emph {et~al.}(2020)\citenamefont
  {Perlepe}, \citenamefont {Oyarzabal}, \citenamefont {Mailman}, \citenamefont
  {Yquel}, \citenamefont {Platunov}, \citenamefont {Dovgaliuk}, \citenamefont
  {Rouzi{\`e}res}, \citenamefont {N{\'e}grier}, \citenamefont {Mondieig},
  \citenamefont {Suturina} \emph {et~al.}}]{perlepe2020metal}%
  \BibitemOpen
  \bibfield  {author} {\bibinfo {author} {\bibfnamefont {P.}~\bibnamefont
  {Perlepe}}, \bibinfo {author} {\bibfnamefont {I.}~\bibnamefont {Oyarzabal}},
  \bibinfo {author} {\bibfnamefont {A.}~\bibnamefont {Mailman}}, \bibinfo
  {author} {\bibfnamefont {M.}~\bibnamefont {Yquel}}, \bibinfo {author}
  {\bibfnamefont {M.}~\bibnamefont {Platunov}}, \bibinfo {author}
  {\bibfnamefont {I.}~\bibnamefont {Dovgaliuk}}, \bibinfo {author}
  {\bibfnamefont {M.}~\bibnamefont {Rouzi{\`e}res}}, \bibinfo {author}
  {\bibfnamefont {P.}~\bibnamefont {N{\'e}grier}}, \bibinfo {author}
  {\bibfnamefont {D.}~\bibnamefont {Mondieig}}, \bibinfo {author}
  {\bibfnamefont {E.~A.}\ \bibnamefont {Suturina}},  \emph {et~al.},\
  }\href@noop {} {\bibfield  {journal} {\bibinfo  {journal} {Science}\ }\textbf
  {\bibinfo {volume} {370}},\ \bibinfo {pages} {587} (\bibinfo {year}
  {2020})}\BibitemShut {NoStop}%
\end{thebibliography}%

\end{document}